\documentclass[prd,showpacs,preprintnumbers,amsmath,amssymb,nofootinbib]{revtex4}
\usepackage{graphicx}
\usepackage{bm}
\usepackage{epsfig}

\begin{document}
\title{The pion form factor at timelike momentum transfers in a dispersion approach} 
\author{D. Melikhov, O. Nachtmann, and T. Paulus}
\affiliation{Institut f\"ur Theoretische Physik, Universit\"at Heidelberg, 
Philosophenweg 16,  D-69120, Heidelberg, Germany}
\date{\today}
\begin{abstract}
We consider a model with $\rho\pi\pi$, $\rho KK$, and gauge-invariant 
$\rho\gamma$ couplings and obtain the pion form factor $F_\pi$ at timelike momentum 
transfers by resummation of pion and kaon loops. We use a dispersion representation for the loop diagrams 
and analyse ambiguities related to subtraction constants. The resulting representation for
$F_\pi$ is shown to have the form of the conventional vector meson dominance formula 
with one important distinction - the effective $\rho$-meson decay constant $f^{\rm eff}_\rho$
turns out to depend on the momentum transfer squared $q^2$. 
For the electromagnetic pion form factor we include in addition the $\rho-\omega$ mixing effects. 
We apply the representations obtained 
to the analysis of the data on the pion form factors 
from $e^+e^-$ annihilation and $\tau$ decay in the region $0\le q^2\le 1$ GeV$^2$ and extract 
the $\rho^-$, $\rho^0$ and $\omega$ masses 
and coupling constants. 


\end{abstract}
\pacs{12.40.Vv, 13.40.Gp,13.65.+i}
\maketitle

\section{Introduction}
The pion {\it electromagnetic} form factor is defined by 
\begin{eqnarray}
\langle \pi^+(p')|J_\mu|\pi^+(p)\rangle=e(p+p')_\mu F_\pi(q^2),\qquad q=p'-p, 
\end{eqnarray}
where $J_\mu$ is the electromagnetic current, $e=\sqrt{4\pi\alpha_{e.m.}}$. 
The form factor is normalized as $F_\pi(0)=1$.

As function of the complex variable $s=q^2$, 
the form factor $F_\pi(s)$ has a cut in the complex $s$-plane starting at the two-pion 
threshold $s=4 m_\pi^2$ which corresponds to two-pion intermediate states. 
There are also cuts related to $K\bar K$ intermediate states and multimeson states ($4\pi$, etc). 
The form factor in the timelike region $(s>0)$ is  
\begin{eqnarray}
F_\pi(s+i0)=|F_\pi(s)|e^{i\delta(s)}, 
\end{eqnarray}
where $\delta(s)$ is the phase. 
For the theoretical description of the form factor in different regions of momentum transfers
different theoretical approaches are used. 

At large spacelike momentum transfers, $-q^2\to\infty$, perturbative QCD (pQCD) 
gives rigorous predictions for the asymptotic behaviour of the form factor  
\cite{bl}
\begin{eqnarray}
\label{pqcd}
F_\pi(q^2)\sim\frac{8\pi f_\pi\alpha_s(-q^2)}{-q^2}, 
\end{eqnarray}
where $\alpha_s$ is the QCD coupling parameter and $f_\pi=130.7\pm 0.4$ MeV is the 
pion decay constant defined by the relation 
\begin{eqnarray}
\langle 0|\bar d\gamma_\mu\gamma_5 u|\pi^+(p)\rangle=ip_\mu f_\pi.
\end{eqnarray}
As the spacelike momentum transfer becomes smaller, 
the form factor turns out to be the result of the interplay of perturbative
and nonperturbative QCD effects, 
with a strong evidence that nonperturbative QCD effects dominate 
in the region $0\le -q^2\le 10$ GeV$^2$ \cite{amn}. The picture based on the concept 
of constituent quarks which effectively account for nonperturbative dynamics 
has proven to be efficient for the description of the form factor in
this region (see for instance \cite{qm}). 

At large timelike momentum transfers, $s\ge 10\div 20$ GeV$^2$, $F_\pi(s)$ can be obtained from the 
analytic continuation of the pQCD formula (\ref{pqcd}). 
At small timelike momentum transfers the situation is more complicated 
since there dynamical details of the confinement mechanism are crucial. 
Quarks and gluons are no longer the degrees of freedom of QCD leading to a 
simple description of the form factor. 
At timelike momentum transfers we are essentially in the region of hadronic singularities 
and typically one relies on methods based on hadronic degrees of freedom. 
In the region of interest to us here, $q^2=0\div 1.5$ GeV$^2$, the 
lightest pseudoscalar mesons are most important. 

There are many approaches to understand the behaviour of the pion form factor at 
timelike momentum transfers from 0 to 1.5 GeV$^2$.

A time honoured approach is based on the vector meson dominance (VMD) model \cite{vmd}. 
In the simplest version of VMD one assumes just the $\rho$-meson dominance, which leads to  
\begin{eqnarray}
\label{piff_a}
F_\pi(s)=\frac{M_\rho^2}{M_\rho^2-s}. 
\end{eqnarray}
This simple formula works with a good accuracy both for small spacelike momentum transfers 
and timelike momentum transfers below the $\pi\pi$ threshold: 
$-1$ GeV$^2\le s\le 4m_\pi^2$.  
For $s$ near the $\pi\pi$ threshold one should take into account effects of the 
virtual pions. In this region momenta of the intermediate pions are small and a 
consistent description of the form factor is provided by chiral perturbation theory 
(ChPT) \cite{chpt}, the effective theory for QCD at low energies.

For higher $s$, in the region of $\rho$ and $\omega$ resonances,  
a similar rigorous treatment of the form factor is still lacking, 
and one has to rely on model considerations. 

Contributions of the two-pion intermediate states may be consistently 
described by dispersion representations. The application of dispersion relations 
has led to the famous Gounaris-Sakurai (GS) formula \cite{gs} which takes into account 
$\rho$-meson finite width corrections due to the virtual pions
\begin{eqnarray}
\label{piff_b}
F_\pi(s)=\frac{M_\rho^2-B_{\rho\rho}(0)}{M_\rho^2-s-B_{\rho\rho}(s)}. 
\end{eqnarray}
The function $B_{\rho\rho}(s)$ corresponds to the two-pion loop diagram. 
The corresponding Feynman integral is linearly divergent, but its imaginary part is defined 
in a unique way. The real part is then reconstructed by a doubly-subtracted 
dispersion representation. The Gounaris-Sakurai prescription 
to fix the subtraction constants reads 
\begin{eqnarray}
{\rm Re}\; B_{\rho\rho}(s)|_{s=M_\rho^2}=0,\qquad 
\frac{d}{ds}{\rm Re}\; B_{\rho\rho}(s)|_{s=M_\rho^2}=0. 
\end{eqnarray}
The phase of the form factor 
\begin{eqnarray}
\label{phase}
{\rm tan}\; \delta (s)=\frac{{\rm Im}B_{\rho\rho}(s)}{M_\rho^2-s-{\rm Re} B_{\rho\rho}(s)}. 
\end{eqnarray}
for the GS prescription agrees well with the experimental data in the region 
$4m_\pi^2<s<0.9$ GeV$^2$. 
But (\ref{piff_b}) gives too small a value (by $\simeq 15\%$) for $|F_\pi(s)|$ at 
$s$ around $M_\rho^2$. 

On the other hand, one can consider a simple VMD ansatz taking only the $\rho$-meson contribution
into account. This should be a good approximation in the region $s=0.5\div 0.8$ GeV$^2$,  
except for the narrow interval near $s\simeq M_\omega^2$  where the $\rho-\omega$ mixing effects 
are important \cite{lefr}. 
The simple VMD ansatz is then very similar to (\ref{piff_b}), but with the numerator replaced 
by the $\gamma\to\rho\to\pi\pi$ transition matrix element: 
\begin{eqnarray}
\label{piff_c}
F_\pi(s)=\frac{\frac{1}{2}g_{\rho\to\pi\pi}f_\rho\;M_\rho}{M_\rho^2-s-B_{\rho\rho}(s)}. 
\end{eqnarray}
Here $g_{\rho\pi\pi}$ and $f_\rho$ are defined according to  
\begin{eqnarray}
\langle \pi(k_1)\pi(k_2)|T|\rho(\varepsilon,k)\rangle&=&
\frac{1}{2}g_{\rho\to \pi\pi} \;\varepsilon_\mu\cdot (k_1-k_2)^\mu,   
\\
\langle 0|J_\mu|\rho^0(\varepsilon,k)\rangle&=&f_\rho M_\rho \varepsilon_\mu, 
\end{eqnarray}
where $\varepsilon_\mu$ is the $\rho$-meson polarization 
and $k$ is the 4-momentum vector. 
Now $|F_\pi(s)|$ from (\ref{piff_c}) describes well the data 
for $s\simeq M_\rho^2$. But extrapolating the Eq. (\ref{piff_c}) to $s=0$ gives 
$F_\pi(0)\simeq 1.15$ in gross violation of the normalization condition $F_\pi(0)=1$. 

Thus, neither (\ref{piff_b}) nor (\ref{piff_c}) can describe the form factor for all 
$s=0\div 1.5$ GeV$^2$: namely, (\ref{piff_b}) leads to a too small value of $|F_\pi|$ at $s=M_\rho^2$, 
whereas the form factor given by (\ref{piff_c}) is far above unity at $s=0$. 
There were many attempts to modify the vector meson dominance or to use related approaches in order to 
bring the results on the pion form factor in agreement with the data
(see \cite{weise,vmd-mod,pich} and papers quoted therein). The pion form factor in the region 
$s=0\div 1.5$ GeV$^2$ is one of the main sources for obtaining masses and coupling constants of 
vector mesons. 
However, with different assumptions on the form of the vector-resonance contribution to 
the pion form factor one obtains different values of masses and couplings. 
Therefore a consistent description of the pion form factor in this region in terms of the low-lying
mesons ($\pi, K, \rho, \omega$) is crucial for extracting reliable values of these parameters. 
Interesting results have been obtained by the authors of \cite{weise} who noticed that 
an effective momentum-dependent $\rho\gamma$ coupling appears in the framework of the effective 
Lagrangian approach. This momentum-dependent $\rho\gamma$ coupling considerably 
improves the description of the pion form factor at timelike momentum transfers in the region 
$0< q^2 <1$ GeV$^2$. Our analysis is similar in spirit to the approach of \cite{weise}. 

The goal of our paper is twofold: First, to analyse the pion form factor in a dispersion approach 
and to establish in this way a general form of the resonance contribution to the form factor, paying
special attention to the analysis of the existing ambiguities.  
Second, to apply the results obtained to the extraction of the $\rho$ and $\omega$ masses and 
couplings from the pion form factor data.  

We make use of a dispersion approach to the pion form factor in a model 
with $\rho\pi\pi$, $\rho KK$, $\omega\pi\pi$, and gauge-invariant $\rho-\gamma$, 
$\omega-\gamma$ and $\rho-\omega$ couplings. 
Our approach allows an exact resummation of pion and kaon loops. 
We pay a special attention to the analysis of ambiguities related to subtractions in 
linearly divergent meson loop diagrams and show that they may be reabsorbed in the 
physical masses and the effective couplings in the expression for the pion form factor.  
After taking into account the $\rho-\omega$ mixing effects the pion form factor 
in the range ${s}=0\div 1$ GeV$^2$ is well described both in magnitude and phase
by a formula which is similar to the VMD expressions (\ref{piff_b}) and (\ref{piff_c}) 
but avoids their pitfalls.

\section{\label{section:2}The model} 
Our model makes use of conventional methods of dispersion theory. First we make an ansatz 
for the effective couplings of the pseudoscalar mesons, vector mesons and the photon. 
These couplings are used in essence  only to calculate the absorptive parts of the 
amplitudes. The complete amplitudes are then obtained by dispersion relations and a Dyson
resummation. We want to make clear from the outset that our effective couplings discussed below 
are not to be compared directly to the effective Lagrangian of  ChPT \cite{chpt} 
and resonance theory in the framework of ChPT \cite{chpt2}. We shall see, however, 
that our model,  
used as explained above, respects all the known results from ChPT for the pion form factor.  
Thus our model can be seen as an alternative to the one of \cite{pich} where ChPT results 
are extended to $F_\pi(s)$ in the range $0\le s\le 1.5$ GeV$^2$ using again a resummation scheme. 

In our model pions interact with the $\rho$-mesons and generate in this way the finite $\rho$ meson width.\footnote{
We do not include into consideration direct four-pion couplings. Neglecting of the latter 
goes along the line of the resonance saturation in the ChPT \cite{chpt2} which states that the coupling 
constants of the effective chiral Lagrangian at order $p^4$ are essentially saturated by the 
meson resonance exchange.} 
The $\rho^0$-meson is coupled to the conserved 
vector current of charged pions as follows:   
\begin{eqnarray}
\label{rhopipi}
L_{\rho\pi\pi}=\frac{i}{2}g\left(\pi^\dagger\partial_\mu\pi-\partial_\mu \pi^\dagger\pi\right)\rho^\mu, 
\end{eqnarray}
where $\rho^\mu$ is the conserved vector field describing the $\rho$ meson. 
We denote in this Section $g\equiv g_{\rho\to\pi\pi}$. 
Matching to the one-loop ChPT \cite{chpt} leads to the relation 
\begin{eqnarray}
\label{chpt}
g_{\rho\to\pi\pi}={2M_\rho}/{f_\pi}. 
\end{eqnarray}
The photon is coupled to the charged pion through the usual minimal coupling, 
\begin{eqnarray}
L_{\gamma\pi\pi}=ie(\pi^\dagger\partial_\mu\pi-\partial_\mu \pi^\dagger\pi)A^\mu.
\end{eqnarray}
We also add a direct gauge-invariant $\gamma\rho$ coupling of the form 
\begin{eqnarray}
\label{rhogamma}
L_{\gamma\rho}=-\frac{1}{4}\frac{ef_\rho}{M_\rho}F^{\mu\nu}G^{(\rho)}_{\mu\nu}, 
\end{eqnarray}
where 
$
F_{\mu\nu}=\partial_\mu A_\nu-\partial_\nu A_\mu,\;
G^{(\rho)}_{\mu\nu}=\partial_\mu \rho_\nu-\partial_\nu \rho_\mu.
$
This model is similar to the model of \cite{klz}. 
No $G$-parity violating $\omega\pi\pi$ or direct $\omega\rho$ couplings are 
included at this stage. 
\begin{figure}
\begin{center}
\begin{tabular}{c}
\mbox{\epsfig{file=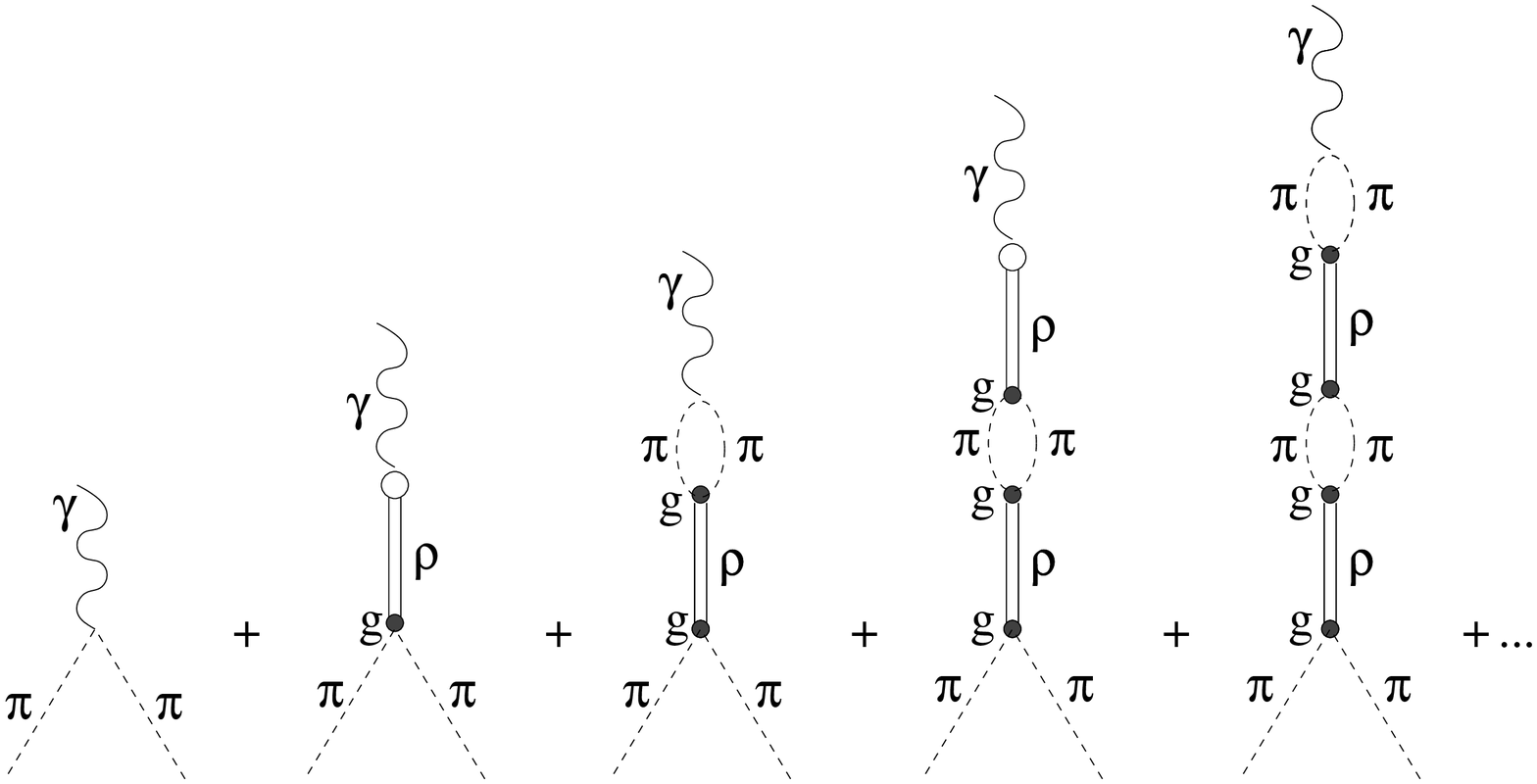,height=4.5cm}}
\end{tabular}
\caption{\label{figure-1:piff}Pion form factor in the picture where 
pions interact via the $\rho$-meson exchange and generate in this way the 
finite $\rho$ meson width. The photon is coupled to the charged pions 
throught the usual minimal coupling, and the direct gauge-invariant 
$\rho-\gamma$ coupling is assumed. No $G$-parity violating effects are included at this stage.}
\end{center}
\end{figure}
As explained above, we calculate the electromagnetic form factor in our model 
by the sum of the diagrams of Fig. \ref{figure-1:piff}. 
Summing all the pion loop insertions, we obtain  
\begin{eqnarray}
\label{fff}
F_\pi(s)=1+\frac{\frac{f_\rho}{2M_\rho}s\cdot g+\frac{g}{2}B_{\rho\gamma}(s)}{M_\rho^2-s-B_{\rho\rho}(s)}
=\frac{M_\rho^2-(1-\frac{f_\rho}{2M_\rho}g)s
+\left\{\frac12g\;B_{\rho\gamma}(s)-B_{\rho\rho}(s)\right\}}{M_\rho^2-s-B_{\rho\rho}(s)}.
\end{eqnarray} 
The quantites $B_{\rho\rho}(s)$ and 
$B_{\gamma\rho}(s)$ correspond to one-loop $\rho\gamma$ and $\rho\rho$ self energy diagrams 
generated by the pion loop. The imaginary parts of these diagrams can be calculated by 
setting the intermediate pions on mass shell. The full functions 
$B_{\rho\rho}(s)$ and $B_{\gamma\rho}(s)$ are constructed from their
imaginary parts by means of the spectral representation with a suitable number
of subtractions and by adding the corresponding subtraction constants.\footnote{This is the usual 
dispersion theory procedure which we adopt since the Feynman integral for the pion one-loop 
diagram leads to a divergent expression.}
For the $\pi\pi$ intermediate states the imaginary parts of the functions 
$B_{\rho\rho}(s)$ and $B_{\gamma\rho}(s)$ satisfy the relations  
\begin{eqnarray}
\label{relationim}
{\rm Im}\; B_{\rho\rho}(s)=g^2 {\rm Im}\; B_{\pi\pi}(s), \qquad 
{\rm Im}\; B_{\rho\gamma}(s)=2\/g\;{\rm Im}\; B_{\pi\pi}(s), 
\end{eqnarray}
where 
\begin{eqnarray}
\label{is}
{\rm Im}\; B_{\pi\pi}(s)\equiv  
I(s,m_\pi^2)=\frac{1}{192\pi}s\left(1-\frac{4m_\pi^2}{s}\right)^{3/2}.  
\end{eqnarray}
For the realistic description we have to take into account also contributions of $K^+K^-$ and $K^0\bar K^0$ 
intermediate states. The coupling constant $g_{\rho\to KK}$ cannot be measured directly. 
We use the relation 
\begin{eqnarray}
\label{gsu3}
2g_{\rho\to KK} =g_{\rho\to \pi\pi}=g,  
\end{eqnarray}
which is valid in the SU(3) limit. Repeating the procedure described above, summing the pion and 
kaon loops we find with (\ref{gsu3})   
\begin{eqnarray}
\label{relationim1}
{\rm Im}\; B_{\rho\rho} &=&g^2 
\left({
{\rm Im} \;B_{\pi\pi} +
\frac{1}{4}\left(
{\rm Im}\;B_{K^+K^-}+{\rm Im}\;B_{K^0\bar K^0}\right)
}\right)
=g^2 \left({
{\rm Im}\; B_{\pi\pi} +
\frac{1}{2}{\rm Im}\;B_{KK}}\right),  
\nonumber \\
{\rm Im}\; B_{\rho\gamma}&=&
2g\left(
{\rm Im}\; B_{\pi\pi} +
\frac{1}{2}{\rm Im}\;B_{K^+K^-}\right)
=2g \left(
{\rm Im} \;B_{\pi\pi} +\frac{1}{2}{\rm Im}\;B_{KK}\right),  
\end{eqnarray}
and hence 
\begin{eqnarray}
\label{rel1}
\frac12\,g\;{\rm Im} \;B_{\rho\gamma}(s)-{\rm Im} \;B_{\rho\rho}(s)=0. 
\end{eqnarray}
It follows from (\ref{rel1}) that the difference $\frac12gB_{\rho\gamma}(s)-B_{\rho\rho}(s)$ 
is a polynomial in $s$ determined by the subtraction conditions. Hence the numerator of 
the pion form factor (\ref{fff}) is also a real polynomial. Therefore, the phase of the form factor is completely determined 
by the denominator. The latter is the usual propagator of the $\rho$ meson with the finite 
width corrections taken into account. 

Let us now consider subtraction constants. 
The function $B_{\gamma\rho}(s)$ describes the coupling of the 
pion to the conserved electromagnetic current. Therefore we must set    
\begin{eqnarray}
\label{bgammarho}
B_{\rho\gamma}(0)=0, 
\end{eqnarray}
such that the charge of the pion remains unrenormalized by higher order corrections.  

The function $B_{\rho\rho}(s)$ determines the behaviour of the $\pi\pi$ elastic 
$J^P=1^-$ partial wave amplitude in which the $\rho$-meson pole is known to be
present in the zero-width limit. Therefore, we require \footnote{This is  
our {\it definition} of the $\rho$ meson mass.  
Notice that various definitions of the mass and the width of 
a resonance are used in the literature. For a different definition see e.g. \cite{anisovich}}
\begin{eqnarray}
{\rm Re}\; B_{\rho\rho}(M_\rho^2)=0. 
\end{eqnarray}
Let us point out that this is essentially our definition of the $\rho$ meson mass. 
One should take into account that various definitions of the mass and the width of 
a resonance

of nonzero width 

Without a loss of generality the second subtraction constant may be fixed by setting  
\begin{eqnarray}
B_{\rho\rho}(s=0)=0. 
\end{eqnarray}
Any other condition would just lead to rescaling of the parameters in the 
formula for the form factor. 

Thus, in a model we are considering when pions interact via a single $\rho$-meson exchange
the general expression for the form factor incorporating 
subtraction ambiguities\footnote{
Assuming more than two subtractions in the 
pion loop diagrams leads to more subtraction constants. This is not 
dictated by the convergence properties of the loop diagrams, 
but is still possible.  
We will not discuss such a case in this paper.} 
in the $\pi\pi$ and $KK$ loop diagrams 
contains three constants $M_\rho^2$, $g$, and $f_\rho$
\begin{eqnarray}
\label{rhodominance}
F_\pi(s)=\frac{M_\rho^2-(1-\frac{f_\rho}{2M_\rho}g)s}
{M_\rho^2-s-B_{\rho\rho}(s)}. 
\end{eqnarray}
Here 
\begin{widetext}
\begin{eqnarray}
B_{\rho\rho}(s)=g^2\;s
\left(
{R(s,m_\pi^2)-R(M^2_{\rho},m_\pi^2)}
+\frac{R(s,m_K^2)-R(M^2_{\rho},m_K^2)}{2}
\right)
+ig^2\;\left(
I(s,m_\pi^2)+\frac{I(s,m_K^2)}{2} \right), 
\end{eqnarray}
$I(s,m^2)$ is defined by (\ref{is}),  and 
\begin{eqnarray}
R(s,m^2)&=&\frac{s}{192\pi^2}\;{\rm V.P.}\;\int_{4m^2}^\infty \frac{ds'}{(s'-s)s'}\left(1-\frac{4m^2}{s'}\right)^{3/2}
\\
\nonumber
&=&\left\{\begin{array}{lll} 
\frac{1}{96\pi^2}
\left(
\frac{1}{3}+\xi^2+\frac{\xi^3}{2}\log\left(\frac{1-\xi}{1+\xi}\right)
\right),  
& \qquad \xi=\sqrt{1-\frac{4m^2}{s}}, 
& \quad {\rm for}\;\; s>4m^2, \\
\frac{1}{96\pi^2}\left(
\frac{1}{3}-\xi^2+\xi^3\,{\rm arctan}\left(\frac{1}{\xi}\right)
\right), 
& \qquad \xi=\sqrt{\frac{4m^2}{s}-1}, 
& \quad {\rm for}\;\; 0<s<4m^2, \\  
\frac{1}{96\pi^2}
\left(
\frac{1}{3}+\xi^2+\frac{\xi^3}{2}\log\left(\frac{\xi-1}{\xi+1}\right)
\right),  
& \qquad \xi=\sqrt{1-\frac{4m^2}{s}}, 
& \quad {\rm for}\;\; s<0, \\
\end{array}\right. 
\end{eqnarray}
where V.P. means the principle value. 
Let us point out that the numerator of the form factor in  (\ref{rhodominance})
is not a constant, but a linear function of $s$. 
This $s$-dependence appears as the direct consequence of the current conservation. 
Eq. (\ref{rhodominance}) can be written in the form of the modified GS formula\footnote{This 
formula is similar to Eq. (5.23) from \cite{weise}. Notice however that our definitions of decay
constants are different from the corresponding definitions of \cite{weise}, therefore 
our formula contains only the physical $\rho$ meson mass, whereas Eq. (5.23) contains both bare and 
physical masses.}
\begin{eqnarray}
\label{piff_our}
F_\pi(s)=\frac{\frac{1}{2}g_{\rho\to\pi\pi}f^{\rm eff}_\rho(s) M_\rho}{M_\rho^2-s-B_{\rho\rho}(s)}. 
\end{eqnarray}
with the effective $s$-dependent $\gamma-\rho$ coupling constant
\begin{eqnarray}
\label{frhoeff}
f^{\rm eff}_\rho(s)=f_\rho\frac{s}{M_\rho^2}+\frac{2(M_\rho^2-s)}{gM_\rho}.   
\end{eqnarray} 
Two remarks are in order here: First, one should be careful with the interpretation of this result: as is clear from  
(\ref{bgammarho}), there is no direct transition of the $\rho$ meson to the {\it real} photon as 
a consequence of the gauge invariant $\rho-\gamma$ coupling \cite{m}. On the other hand, 
the {\it effective} coupling $f^{\rm eff}_\rho(s)$ is clearly nonzero at $s=0$. Therefore the 
pion form factor {\it looks} as the direct $\rho-\gamma$ coupling also for the real photon. 
This is just the usual vector meson dominance. The latter thus emerges as the direct consequence 
of our assumption that the vector meson couples to the same pion current as the photon. 
For further discussions of the relationship between VMD and gauge invariance we refer 
to \cite{klz}. If we use the ChPT relation (\ref{chpt}), which agrees perfectly with the 
measured value of $g_{\rho\to\pi\pi}$, then (\ref{frhoeff}) leads to an interesting relation 
\begin{eqnarray}
f_\rho^{\rm eff}(s=0)=f_\pi.
\end{eqnarray}  
Note that the phase of $F_\pi(s)$ in (\ref{piff_our}) is still given by Eq. (\ref{phase})  
and is completely determined by the function $B_{\rho\rho}(s)$.

Second, it should be noted that the $\rho$-meson contribution to the pion form factor given by 
(\ref{piff_our}) does not fall down sufficienty fast in the limit $|s|\to\infty$. Thus, 
for a realistic description of the pion form factor in a broader range of timelike 
momentum transfers, coupling of pions to higher vector resonances $\rho'$ etc should 
be added. However, if a vector-meson coupling to the photon is described by the 
gauge-invariant expression (\ref{rhogamma}), contributions of higher resonances to $|F_\pi|$ 
are negligible in the region $0\le s\le 1$ GeV$^2$ (see recent analyses \cite{pich}). 
We are interested only in this region and therefore do not take higher resonances $\rho'$ 
etc into consideration. 
\section{\label{section:3}The $\bm\rho-\bm\omega$ mixing}
In Section \ref{section:2} we discussed the $\rho$ contribution to the pion form factor 
(the bare $\rho$ plus the effects of the $\rho$-meson width due to the light-meson loops). 
This analysis is sufficient for describing the pion form factor of the charged vector current using
the CVC relation. For the electromagnetic pion form factor it is necessary to take into account the $\rho-\omega$ 
mixing effects. The $\omega$ is coupled to the pions and the photon similarly to the $\rho^0$-meson 
[cf. (\ref{rhopipi}) and (\ref{rhogamma})]    
\begin{eqnarray}
\label{omegapipi}
L_{\omega\pi\pi}=\frac{i}{2}g_{\omega\to\pi\pi}
\left(\pi^\dagger\partial_\mu\pi-\partial_\mu \pi^\dagger\pi\right)\omega^\mu, \qquad
L_{\gamma\omega}=-\frac{1}{4}\frac{ef_\omega}{M_\omega}F^{\mu\nu}G^{(\omega)}_{\mu\nu},   
\end{eqnarray}
$\omega_\mu$ being a conserved vector field describing the $\omega$-meson and 
$G^{(\omega)}_{\mu\nu}=\partial_\mu \omega_\nu-\partial_\nu \omega_\mu$. 

It has proven useful to classify various contributions to hadronic amplitudes according 
to their formal order in the $1/N_c$ expansion \cite{chpt}, where $N_c$=3 is the number 
of colours in QCD. In the language of the $1/N_c$ expansion the analysis of the previous 
section corresponds to taking into account the leading order $1/N_c$ process 
(resonance in a zero-width approximation) and the subleading $O(1/N_c)$ effects of 
the meson loops.\footnote{
Recall that pion and kaon loop diagrams are of order $1/N_c$ and of order $p^4$ of the momentum expansion.}
Performing a resummation of these meson loops gave our dispersion description of the form factor. 

A corresponding treatment of the $\rho-\omega$ mixing effects then requires taking into account 
the leading and subleading $1/N_c$ effects as well. 
To leading order in $ 1/N_c $, meson loops do not contribute and therefore the only effect is the direct 
$\rho-\omega$ transition described in terms of the direct coupling (Fig. 2). 

At subleading $1/N_c$ order  several meson loop diagrams shown in Fig. 2 emerge. 
We make use of spectral representations for loop diagrams, i.e. we calculate directly
the imaginary parts and then reconstruct the full function by means of the spectral 
integral with a relevant number of subtractions. Subtraction constants are then either fixed by 
physical constraints or determined by the experimental data. 
Let us point out an important feature related to our dispersion calculation: 
the direct $\rho-\omega$ coupling (a leading $1/N_c$ process) 
and the {\it real part} of the $\rho-\omega$ 
mixing loop diagrams at $q^2=M_\rho^2$   
(a subleading $1/N_c$ process) contribute to the form factor precisely the same way, such that 
only their sum has a physical meaning. 
We therefore account for the net effect of these two contributions by a single subtraction constant 
and do not consider the direct $\rho-\omega$ coupling separately. 

We have analysed in Section \ref{section:2} the $\rho$-meson self-energy function $B_{\rho\rho}$ which determines the propagator of
the interacting $\rho$ meson. Let us now discuss a similar self-energy function of the $\omega$-meson 
$B_{\omega\omega}$ and the off-diagonal $\rho-\omega$ function $B_{\rho\omega}$ which describes the 
$\rho-\omega$ mixing. 

The function $B_{\omega\omega}$ determines the $\omega$-propagator 
$D_\omega(s)=1/(M_\omega^2-s-B_{\omega\omega})$ 
in the absence of the $\rho-\omega$ mixing effects. The main contribution to ${\rm Im}\;B_{\omega\omega}$ 
is given by the three-pion intermediate states. This ${\rm Im}\;B_{\omega\omega}$ should then be
inserted into a dispersion integral to get $B_{\omega\omega}$. 
However, because of the small width of the
$\omega$-resonance,  it is sufficient for our analysis to consider as 
a simple ansatz a constant
$B_{\omega\omega}$ 
\begin{eqnarray}
\label{bomegaomega}
B_{\omega\omega}=i\Gamma^{tot}_\omega M_\omega. 
\end{eqnarray} 
Possible processes which contribute to the $\rho-\omega$ mixing amplitude $B_{\rho\omega}=B_{\omega\rho}$ 
are shown in Fig. \ref{fig:mixing1}. 
The coupling contants which determine the relative 
strength of the diagrams in Fig. \ref{fig:mixing1} are shown in Table \ref{table:pdg}.  
\begin{figure*}
\begin{center}
\mbox{\epsfig{file=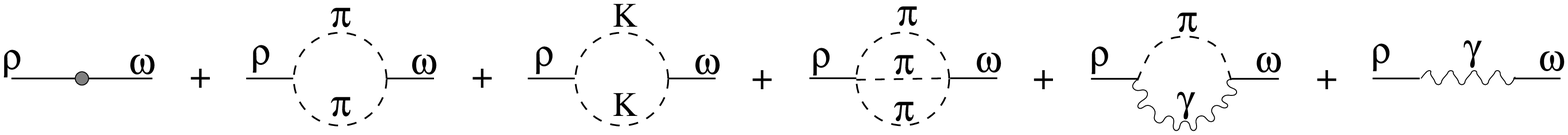,width=15cm}}
\caption{\label{fig:mixing1}Diagrams which contribute 
to the $\rho-\omega$ mixing amplitude $B_{\rho\omega}$. 
The direct $\rho-\omega$ mixing diagram is the only diagram which emerges to 
leading order in $1/N_c$; meson-loop diagrams are subleading $1/N_c$ effect.}
\end{center}
\end{figure*}
\begin{table*}
\caption{\label{table:pdg}
Masses and rates for vector mesons from PDG \protect\cite{pdg} and the corresponding 
decay constants. Recall the SU(2)-limit relations $f_\rho=3f_\omega$.}
\centering
\vspace{.5cm}
\begin{tabular}{|c|c|c|c|c|c||c|c|}
\hline
Res. &  $M$, MeV & $\Gamma^{\rm tot}$, MeV &  $\Gamma_{e^+e^-}$, keV & 
  $Br(\pi^+\pi^-)$ &   $Br({\pi^0\gamma})$ & $f_V$, MeV 
& $g_{V\to 2\pi}$  \\
\hline
$\rho^0$ &  769.0$\pm$0.9    &  150.7$\pm$ 2.9  & 6.77$\pm$ 0.32 & 100\% 
&  $(6.8\pm 1.7)10^{-4}$ 
&  152$\pm$5  
&  11.8$\pm$0.2  \\ 
$\omega$ &  782.57$\pm$0.12  &  8.44$\pm$ 0.09  & 0.60$\pm$ 0.02 & (2.21$\pm$ 0.3)\% 
& $(8.5\pm 0.5)10^{-2}$
&   45.3$\pm$ 0.9  
&   0.4$\pm$0.02  \\  
\hline
\end{tabular}
\end{table*}
One finds (see also Ref. \cite{gourdin}) that the main contribution 
to the imaginary part of the $\rho-\omega$ mixing amplitude $B_{\rho\omega}$ 
is given by the diagrams with two-pion and two-kaon intermediate states. 
To obtain the full $B_{\rho\omega}$ we write again a dispersion representation  
with two subtractions. 
The imaginary parts of these diagrams can be calculated in analogy to (\ref{relationim}) in terms of the coupling constants 
$g_{V\to PP}$ ($V=\rho,\omega$, $P=\pi,K$) defined according to the relation 
\begin{eqnarray}
\langle P(k_1)\bar P(k_2)|T|V(\varepsilon,k) \rangle&=&\frac{1}{2}g_{V\to PP} \varepsilon_\mu^{(V)}(k_1-k_2)^\mu. 
\nonumber
\end{eqnarray} 
For instance, the imaginary part of the diagram with the $\pi\pi$ intermediate state
is equal to $g_{\rho\to\pi\pi}g_{\omega\to\pi\pi}I(s,m_\pi^2)$. 

The same arguments as used to show the relation (\ref{rel1}) between ${\rm Im}\; B_{\rho\gamma}$ 
and ${\rm Im}\; B_{\rho\rho}$ lead to 
\begin{eqnarray}
\label{rel2}
g_{\rho\to\pi\pi} {\rm Im}\;B_{\rho\omega}(s)-g_{\omega\to\pi\pi} {\rm Im} \;B_{\rho\rho}(s)=0. 
\end{eqnarray}
Hence, the combination $g_{\omega\to\pi\pi}  B_{\rho\rho}-g_{\rho\to\pi\pi}  B_{\rho\omega}$ is a 
polynomial in $s$. The $\rho-\omega$ mixing effects are sizeable only in the
narrow vicinity of $s=M_\omega^2$, so we may set  
\begin{eqnarray}
\label{Delta}
g_{\rho\to\pi\pi} B_{\rho\omega}-g_{\omega\to\pi\pi} B_{\rho\rho}=s\;\Delta, 
\end{eqnarray}
and the value of $\Delta$ will be found from the fit to the pion form factor. 
As we have explained above, the {\it real} part of the function $B_{\rho\omega}$ at $s\simeq M_{\rho,\omega}^2$ 
includes the direct $\rho-\omega$ coupling. 

\section{The pion electromagnetic form factor with the $\bm\rho-\bm\omega$ mixing effects} 
In the problem of the $\rho-\omega$ mixing, the constant $g_{\omega\to 2\pi}$
is a natural small parameter, and the expansion of the pion form factor 
in powers of this parameter can be constructed. We can safely neglect all terms 
of order $O(g_{\omega\to\pi\pi}^2)$ and limit ourselves 
to the first order analysis.
The diagrams which describe the contributions to the form factor of first order in $g_{\omega\to 2\pi}$ 
are shown in Fig. \ref{fig:omegarhomixing}.
\begin{figure}[b]
\begin{center}
\mbox{\epsfig{file=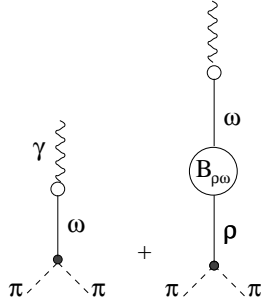,height=4cm}}
\caption{\label{fig:omegarhomixing} Diagrams for the pion form factor which emerge at first 
order of the expansion in $g_{\omega\to \pi\pi}$. 
In this figure the $\rho$ and $\omega$ propagators are   
$D_\rho=1/(M_\rho^2-s-B_{\rho\rho})$ and $D_\omega=1/(M_\omega^2-s-B_{\omega\omega})$, 
respectively.}
\end{center}
\end{figure}
Adding the corresponding expressions to the result (\ref{piff_our}) we get 
for the pion form factor  
\begin{eqnarray}
\label{pifffinal}
F_\pi(s)=\frac{\frac{1}{2}g_{\rho\to\pi\pi}f^{\rm eff}_\rho(s) M_\rho}{M_\rho^2-s-B_{\rho\rho}(s)}
+\frac{\frac{1}{2}g_{\omega\to\pi\pi}\frac{f_\omega s}{M_\omega}}{M_\omega^2-s-B_{\omega\omega}(s)} 
\left\{ 
\frac{M_\rho^2-s+\Delta\cdot s}{M_\rho^2-s-B_{\rho\rho}(s)}
\right\}+O(g^2_{\omega\to\pi\pi}). 
\end{eqnarray}
We use this expression for the numerical analysis of the data for the pion electromagnetic 
form factor in the next Section.  

\end{widetext}

\section{Numerical analysis}
In this Section we apply the formulas obtained to the analysis of the data on the electromagnetic 
and charged-current pion form factors and extract in this way the resonance masses and coupling constants. 
We include the contributions of the $\rho(770)$ and $\omega(782)$ resonances and neglect the 
higher vector resonances $\rho(1450)$ and $\rho(1700)$ (For a discussion of these latter see \cite{dosch}). 
As can be seen from the analysis of \cite{aleph}, the influence
of the latter upon the pion form factor is negligible in the region $s\le 1000$ MeV. We therefore 
extract the $\rho$ and $\omega$ parameters making use of the form factor data for $s\le 1000$ MeV.

\subsection{The pion electromagnetic form factor}
We fit the available data on the phase \cite{phase} and the modulus \cite{data,data1} of the pion 
electromagnetic form factor to  
(\ref{pifffinal}) which includes the $\rho-\omega$ mixing effects. 
The parameters extracted from the best fit to the {\it phase} and the {\it modulus} of the form 
factor, separately, are shown in Tables \ref{table:fittingphase} and \ref{table:fittingmodulus}, 
respectively. 
The form factor turns out to be weakly sensitive to $g_{\omega\to\pi\pi}$ and $f_\omega$ for 
which we use the values from Table \ref{table:pdg}. 
\begin{table*}[ht]
\caption{\label{table:fittingphase}
The upper limit of the $\sqrt{s}$-range of the data from \cite{phase} used for fitting the {\it phase} of the pion form 
factor and the corresponding fitted parameters $M_\rho$ and $g_{\rho\to2\pi}$. Error estimates 
as given by the FUMILI program are shown.}
\centering
\begin{tabular}{|c||l|l|l|l|l|}
\hline
$Q_{\rm upper}$, MeV         & 710 (5 pts)    & 775 (10 pts)   &  850 (15 pts)	 & 965  (20 pts)  \\
\hline
$M_{\rho^0}$, MeV	     & 772.7$\pm$1.3  & 773.4 $\pm$0.8 &  773.0$\pm$0.6  & 771.1$\pm$0.6  \\
\hline
$g_{\rho^0\to\pi^+\pi^-}$    & 12.05$\pm$0.07 & 12.0 $\pm$0.05 &  12.0 $\pm$0.04 & 11.87$\pm$0.04 \\
\hline
\end{tabular}
\end{table*}
The extracted resonance parameters turn out to be rather sensitive to the upper limit 
$\sqrt{s}\le Q_{\rm upper}$ of 
the data points included into the fit procedure. The extracted masses and couplings depending on $Q_{\rm upper}$ 
are shown in Table \ref{table:fittingphase} and \ref{table:fittingmodulus}. This dependence might signal that 
the errors in the extracted masses and coupling constants are in fact sizeably greater than those quoted 
in PDG \cite{pdg}. Obviously, the error estimates provided by the popular FUMILI program should be taken with
some care. 

Our best estimates for the $\rho$ and $\omega$ parameters from a combination of the fits to the phase and 
the modulus are presented in Table \ref{table:fitresults} in Section \ref{section:6}. 
We arrive at these values as follows: The parameter values from the last columns of Tables \ref{table:fittingphase}
and \ref{table:fittingmodulus} should be the most reliable ones, since they correspond to the biggest data sets. 
On the other hand, the errros given by the FUMILI program cannot be trusted. We took the average of the values for  
$M_\rho^0$ and $g_{\rho\to\pi\pi}$, weighting the values from the modulus fit by a factor 2/3 and those from the
phase fit by 1/3. The errors in Table \ref{table:fitresults} are our educated guesses. 

\begin{figure}
\begin{center}
\begin{tabular}{c}
\mbox{\epsfig{file=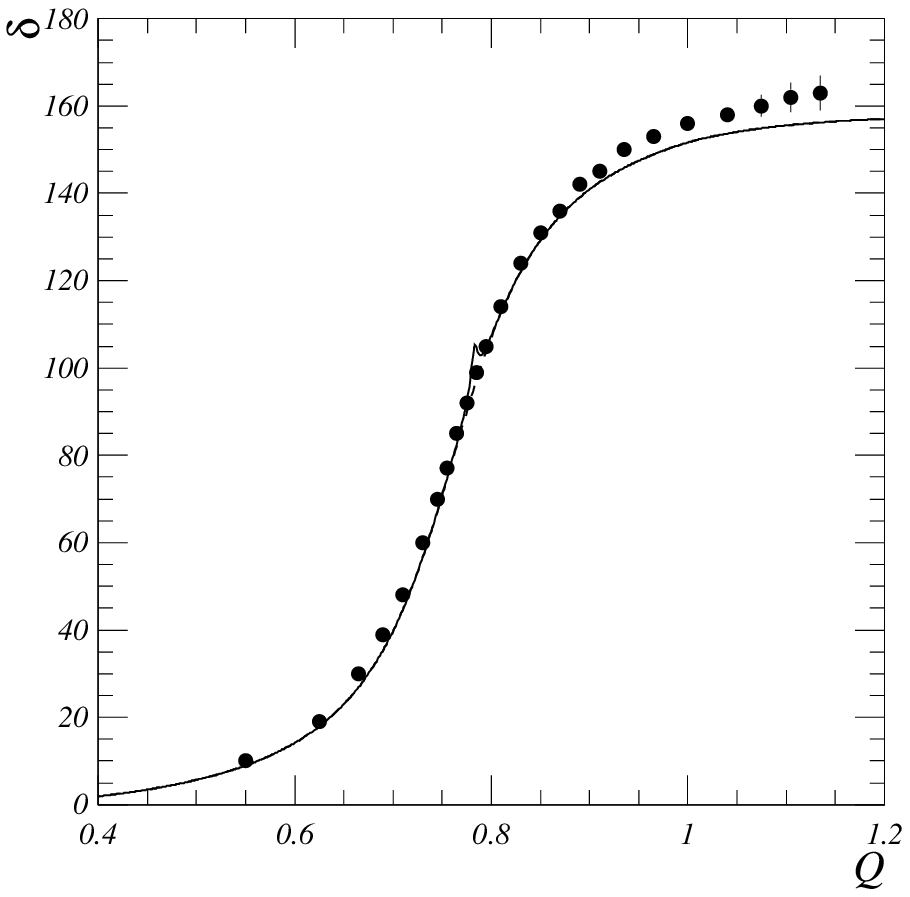,width=7cm}}\\
     \mbox{\epsfig{file=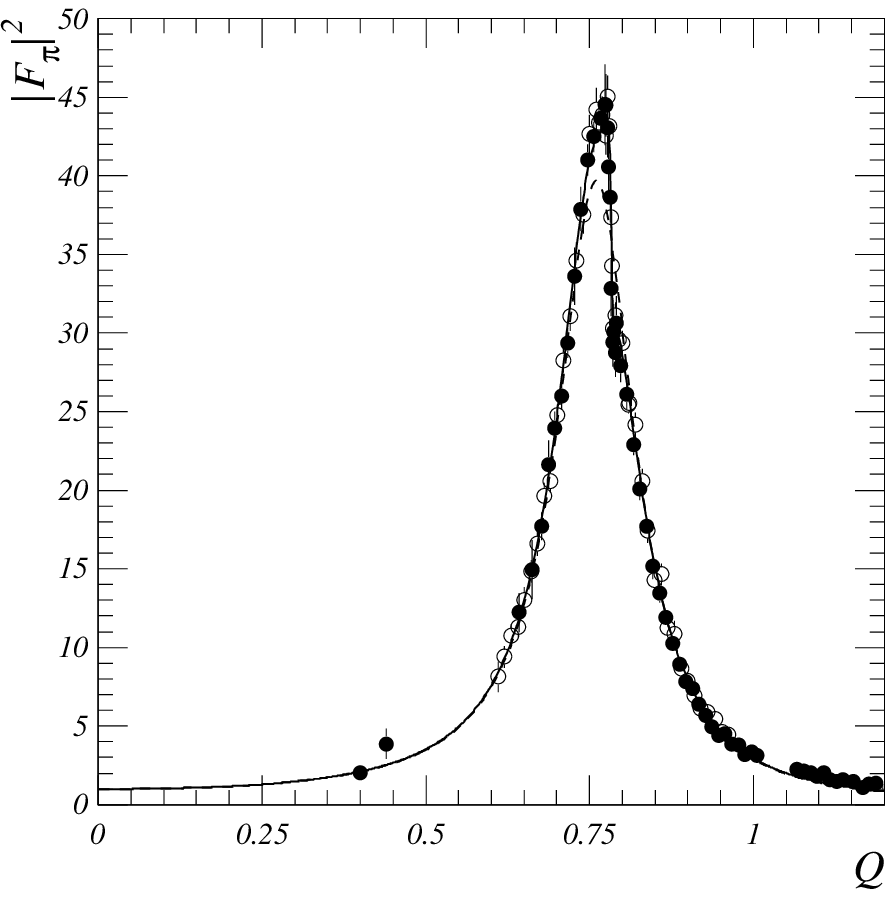,width=7cm}}\\
    \mbox{\epsfig{file=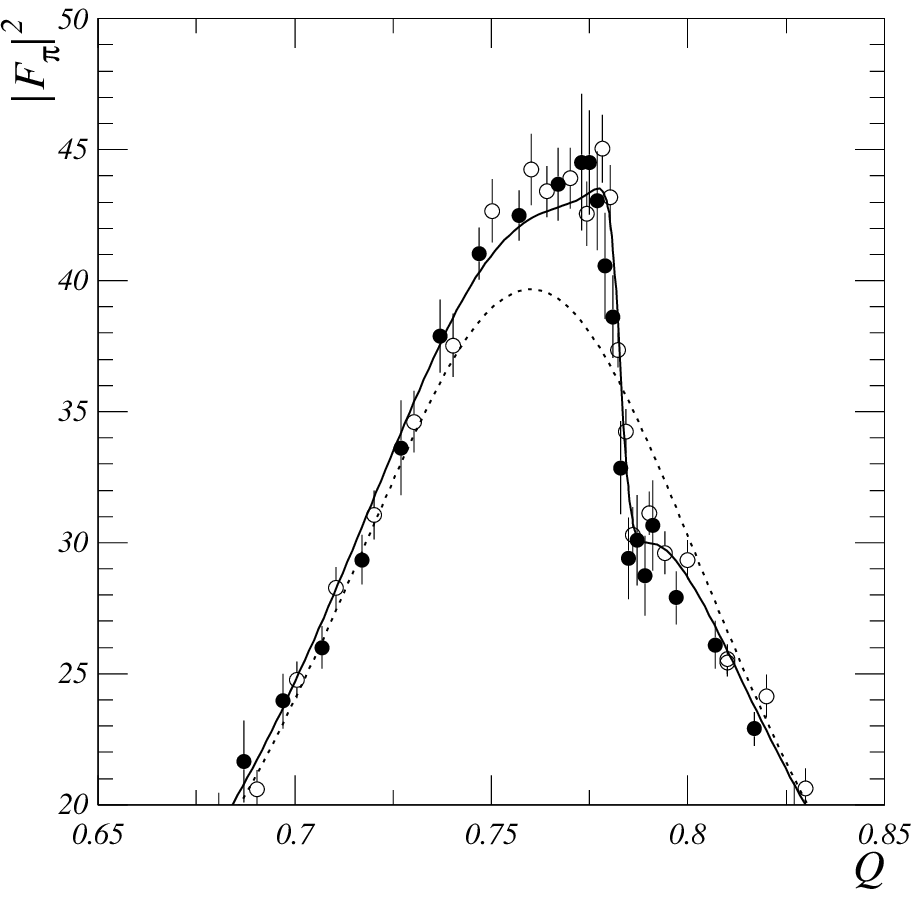,width=7cm}}
\end{tabular}
\caption{\label{fig:piff}Pion form factor vs $Q=\sqrt{s}$ (GeV), calculated for the central values of the 
parameters from Table \ref{table:fitresults}. 
Solid line - the form factor with the full $\rho-\omega$ mixing effects. 
Dotted line - the contribution of $\rho$. 
Data on the phase from \cite{phase}; data on the modulus from \cite{data} (solid) and \cite{data1} (empty).}
\end{center}
\end{figure}

The pion elastic form factor calculated with the central values of the parameters from 
Table \ref{table:fitresults} is shown in Fig. \ref{fig:piff}. 
Both the phase and the magnitude of the form factor are well described, 
except for the phase at $\sqrt{s} > 0.9$ GeV.  
\begin{table*}[hb]
\caption{\label{table:fittingmodulus}
The upper limit of the $Q$-range of the data \cite{data}, used for fitting the {\it modulus} of the pion form 
factor and the corresponding fitted parameters $M_\rho$, $f_\rho$, $g_{\rho\to2\pi}$, 
$M_\omega$, and $\Delta$. The last column 
shows the result of the fit to the combined data on $|F_\pi|$ from \cite{data} and \cite{data1}.  
Error estimates as given by the FUMILI program are shown.}
\centering
\vspace{.1cm}
\begin{tabular}{|c||l|l|l|c|c|}
\hline
$Q_{\rm upper}$, MeV       & 820 (27 pts)  & 950 (40 pts)     &  1000 (45 pts)   & 960  (40 pts \cite{data} +
45 pts \cite{data1})  \\
\hline
$M_{\rho^0}$, MeV          & 774.7$\pm$0.3 & 776.1 $\pm$0.2   &  773.6$\pm$0.2   &  775.5$\pm$0.1   \\
\hline
$f_{\rho^0}$, MeV          & 147.7$\pm$0.2 & 148.2 $\pm$0.1   &  149.0$\pm$0.1   &  149.4$\pm$0.1   \\
\hline
$g_{\rho^0\to\pi^+\pi^-}$  & 11.37$\pm$0.03 & 11.38$\pm$0.01  &  11.7$\pm$0.01   &  11.5 $\pm$0.05  \\
\hline
$M_{\omega}$, MeV          & 782.5$\pm$0.3  & 781.3$\pm$0.2   &  781.9$\pm$0.2   &  782.5$\pm$0.2   \\
\hline
$\Delta$                   & 0.180$\pm$0.007& 0.191$\pm$0.006 &  0.183$\pm$0.006 &  0.170$\pm$0.007 \\
\hline
\end{tabular}
\end{table*}

\newpage
\subsection{The pion charged current form factor} 
\begin{figure}[hb]
\begin{center}
\mbox{\epsfig{file=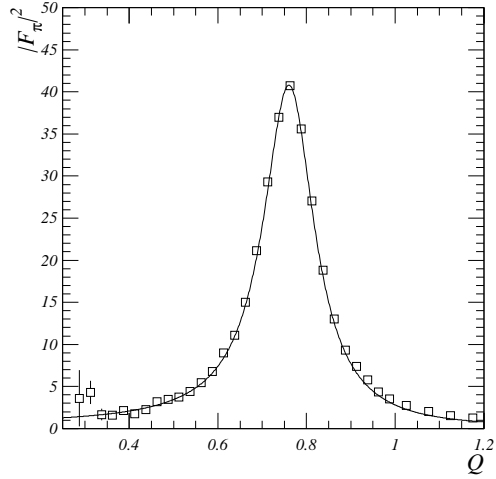,width=7cm}}\\
\caption{\label{fig:piff-tau}The charged current pion form factor   
calculated for the parameter set obtained for $Q_{\rm upper}=900$ MeV 
from Table \ref{table:fittingmodulustau} and the 
CLEO data \cite{cleo}.}
\end{center}
\end{figure}

The amplitude of the weak tranistion $\tau^-\to \pi^-\pi^0 \nu_\tau$  
can be parametrized in terms of the two $\pi^-\to \pi^0$ transition form factors as follows 
\begin{eqnarray}
\langle \pi^0(p')|\bar u\gamma_\mu d|\pi^-(p)\rangle=
\frac{1}{\sqrt2}F_\pi^+(q^2) 
(p'+p)_\mu + \frac{1}{\sqrt2}F_\pi^-(q^2)q_\mu.
\end{eqnarray}
In the isospin limit $F_\pi^-=0$ and $F_\pi^+=F_\pi$. 
These relations should work well for all $q^2$ except for the region of the $\rho$ 
and $\omega$ resonances: The form factor $F_\pi$ contains contributions of the $\rho^0$ and 
$\omega$-resonance, whereas the contribution analogous to $\omega$ is absent in $F_\pi^+$. 
Thus, the charged current form factor $F_\pi^+$ as measured in the $\tau^-\to \pi^0\pi^-\nu_{\tau}$ decay 
is given in our model by the the modified $\rho$-dominance formula (\ref{piff_our}). 
Comparison with the ALEPH \cite{aleph} and CLEO \cite{cleo}  data 
allows the extraction of the masses and coupling constants of the $\rho^-$. 
We give the correponding numbers in Table \ref{table:fittingmodulustau} and 
plot the form factor in Fig. \ref{fig:piff-tau}. 
\begin{table*}[hb]
\caption{\label{table:fittingmodulustau}
Fit to the pion charged current form factor from the CLEO data on the  
$\tau^-\to \pi^-\pi^0 \nu_\tau$ decay. The upper limit $Q_{\rm upper}$ of the $\sqrt{s}$-range of the data used 
and the corresponding fitted parameters for the $\rho^-$ meson.  
Error estimates as given by the FUMILI program are shown.}
\centering
\vspace{.5cm}
\begin{tabular}{|c||c|c|c|c|c|}
\hline
$Q_{\rm upper}$, MeV       & 760 (18 pts)      &  900 (23 pts)      & 1025  (28 pts)	\\
\hline
$M_{\rho^-}$, MeV          & 768.8 $\pm$ 0.3   &  775.1 $\pm$ 0.1   &  776.9 $\pm$ 0.1  \\
\hline 
$f_{\rho^-}$, MeV          & 144.9 $\pm$ 0.3   &  150.3 $\pm$ 0.1   &  150.1$\pm$ 0.1	\\
\hline
$g_{\rho^-\to\pi^0\pi^-}$  & 11.22 $\pm$ 0.02  &  11.34 $\pm$ 0.01  &  11.80 $\pm$ 0.05 \\
\hline
\end{tabular}
\end{table*}

\newpage
\section{\label{section:6}Discussion and summary} 
\begin{figure}[hb]
\begin{center}
\mbox{\epsfig{file=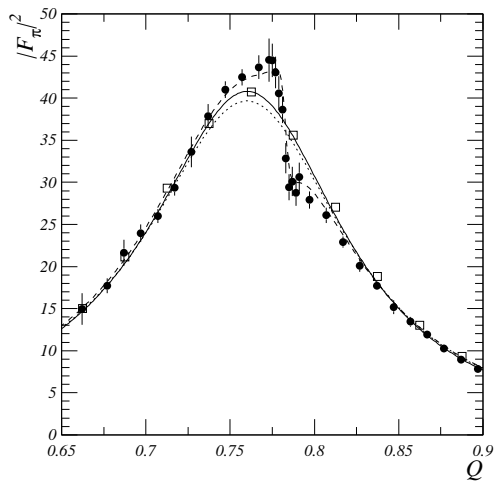,width=7cm}}\\
\caption{\label{fig:piff-both}Comparison of the electromagnetic pion form factor (full dots) 
\cite{data,data1} 
and the charged current form factor from the $\tau^-\to\pi^-\pi^0\nu_\tau$ decay (squares) 
\cite{cleo} with our fits. The fits to the electromagnetic pion form factor show the sum of the $\rho$ and 
$\omega^0$ contributions (dashed) and 
the $\rho^0$ contribution (dotted).  
The fit to the charged current pion form factor is the solid line.}
\end{center}
\end{figure}

We analysed the pion electromagnetic and charged current form factors at timelike momentum transfers 
in a dispersion approach. 
Our results are as follows: 
\begin{itemize}
\item[1.]
We considered a model with $\rho\pi\pi$, $\rho KK$, $\omega\pi\pi$, $\omega KK$ and  
gauge-inavariant $\rho\gamma$ and $\omega\gamma$ couplings. 
The pion form factor is obtained by 
a resummation of pion and kaon loops leading to the finite width of the $\rho$-meson.  
The imaginary parts of the loop diagrams were calculated in terms of the 
$g_{\rho\to\pi\pi}$ coupling, and the real parts were constructed by 
dispersion representations requiring two subtractions. The  subtraction constants 
were related to the observed values of $M_\rho$, $f_\rho$ and $g_{\rho\to\pi\pi}$. 
The resulting expression for the pion form factor takes the form of the 
vector meson dominance formula with one important distinction: the effective decay constant 
$f_\rho^{\rm eff}$ depends linearly on $s$, the momentum transfer squared.  

We have also taken into account the $\rho-\omega$ mixing in the electromagnetic pion form factor.  

\item[2.]
The values of $M_{\rho^0}$, $f_{\rho^0}$, $g_{\rho^0\to\pi^+\pi^-}$, 
$M_\omega$, and the $\rho-\omega$ mixing parameter were extracted from the fit to the pion electromagnetic 
form factor at $\sqrt{s}=0\div 1.0$ GeV (see  Fig. \ref{fig:piff}) where contributions of higher vector resonances 
are negligible. 
The $\rho-\omega$ mixing 
was found to give a sizeable contribution to the electromagnetic form factor in the region 
$\sqrt{s}= 0.74\div 0.82$ GeV: it leads to the increase of $|F_\pi|^2$ 
by 10\% at $s=M_\rho^2$ and by almost 30\% at $s=M_\omega^2$. 

The values $M_{\rho^-}$, $f_{\rho^-}$, $g_{\rho^-\to\pi^0\pi^-}$ were obtained by the fit  
to the pion charged current form factor measured in the $\tau^-\to\pi^-\pi^0\nu_{\tau}$ decay,   
(see Fig. \ref{fig:piff-tau}). The corresponding numbers are presented in Table \ref{table:fittingmodulustau}. 
Our estimate for the central value of the $\rho^-$ mass is given in Table \ref{table:fitresults}. 
Let us point out that our fitted value for $g_{\rho\to 2\pi}$ 
agrees perfectly with the ChPT prediction $g_{\rho\to 2\pi}=2M_\rho/f_\pi$=11.7.

Our best estimates for the $\rho$ and $\omega$ parameters are presented in Table \ref{table:fitresults}.
The masses, the weak decay constants and the pionic coupling constants of the neutral and charged 
$\rho$-mesons were found to be equal within the errors.
We notice that our central values of the $\rho$-masses are 2-3 MeV higher than the corresponding 
numbers obtained from the same reactions by PDG \cite{pdg}. 
Fig. \ref{fig:piff-both} presents comparison of the data and the theoretical curves for 
the electromagnetic and charged current pion form factors.  
\end{itemize}
\vspace{-.3cm}
\begin{table}[ht]
\caption{\label{table:fitresults}
Masses and  decay constants of vector mesons and the $\rho-\omega$ 
mixing parameter $\Delta$ [see (\ref{Delta})] as obtained by our analysis.}   
\centering
\vspace{.2cm}
\begin{tabular}{|c|c|c|c|c|c|}
\hline
$M_{\rho^-}$, MeV &
$M_{\rho^0}$, MeV &
$M_\omega$, MeV &
$f_{\rho}$, MeV & 
$g_{\rho\to\pi\pi}$ &
$\Delta$ \\
\hline
  775$\pm$ 2    &         
  774$\pm$ 2    & 
  782.0$\pm$0.5 &  
  149$\pm$ 1    &    
11.6$\pm$ 0.3   & 
0.17$\pm$ 0.02  \\ 
\hline
\end{tabular}
\end{table}

\newpage
\section{Acknowledgments}
We are grateful to V.~Anisovich, A.~Donnachie, M.~Jamin and E.~de Rafael for useful discussions  
and to J. Urheim for correspondence on the CLEO data. 
The work was supported by the Alexander von Humboldt-Stiftung and 
the German Bundesministerium f\"ur Bildung und Forschung under project 05 HT 9 HVA3. 


\end{document}